\documentclass[aps,pre,twocolumn,groupedaddress,floatfix,longbibliography,superscriptaddress]{revtex4-1}
\usepackage{amsmath}
\usepackage{amssymb}
\usepackage{amsfonts}
\usepackage{graphicx}
\usepackage{bbold}
\usepackage{bm}
\usepackage{bm}
\usepackage{times,float}
\usepackage{graphicx}
\usepackage[usenames,dvipsnames,svgnames]{xcolor}
\usepackage{hyperref}
\hypersetup{colorlinks=true, linkcolor=Blue, citecolor=Green,urlcolor=Blue}

\newcommand{\be}{\begin{equation}}
\newcommand{\ee}{\end{equation}}
\newcommand{\bea}{\begin{eqnarray}}
\newcommand{\eea}{\end{eqnarray}}
\newcommand{\nn}{\nonumber}
\newcommand{\ii}{\mathrm{i}}

\newcommand{\Z}{\mathcal{Z}}

\begin{document}
\title{Volume effects on the QCD critical end point from thermal fluctuations within the super statistics framework}
\author{Jorge David Casta\~no-Yepes}
\email{jcastano@uc.cl}
\affiliation{Instituto de Física, Pontificia Universidad Católica de Chile, Vicuña Mackenna 4860, Santiago, Chile.}
\author{Fernando Martínez Paniagua}
\affiliation{Ingenier\'ia F\'isica, Facultad de Ingenier\'ia, Universidad Aut\'onoma de Quer\'etaro, C.P. 76010 Quer\'etaro, Qro., Mexico.}
\author{Victor Muñoz-Vitelly}
\affiliation{Instituto de Ciencias
  Nucleares, Universidad Nacional Aut\'onoma de M\'exico, Apartado
  Postal 70-543, M\'exico Distrito Federal 04510,
  Mexico.}
\author{Cristian Felipe Ramirez-Gutierrez}
\affiliation{Ingenier\'ia F\'isica, Facultad de Ingenier\'ia, Universidad Aut\'onoma de Quer\'etaro, C.P. 76010 Quer\'etaro, Qro., Mexico.}
\affiliation{Universidad Polit\'ecnica de Quer\'etaro, El Marqu\'es,   76240 Quer\'etaro, Mexico.}
%
\begin{abstract}
{We investigate the impact of the finite volume and the thermal fluctuations on the Critical End Point of the QCD phase diagram. To do so, we implement the super statistics framework with Gamma, $F$, and log-normal distributions and their relation with the Tsallis non-extensive thermodynamics. We compute an effective thermodynamic potential as a function of the inverse temperature fluctuations and explicit dependence on the system volume. To find an analytic expression for the effective potential, we expand the modified Boltzmann factor by using the equilibrium thermodynamic potential computed in the Linear Sigma Model coupled to quarks. We find that the pseudo-critical temperature of transition at vanishing baryon chemical potential is modified by the size of the system being about $7\%$ lower for small volumes. Additionally, the critical endpoint moves to higher densities and lower temperatures (about $12\%$ in both cases). Interestingly, the results are quantitatively the same when the parameter that models the out-of-equilibrium situation is modified, indicating that the chiral symmetry restoration is robust against the thermal fluctuations in this approximation.}
\end{abstract}
\pacs{xxx xxx xx xx}
\maketitle
\section{Introduction}\label{sec:intro}
{
In the understanding of the strongly interacting matter under extreme conditions, the study of the QCD-phase diagram has become of great interest in the last few years. In particular, determining the location of the Critical End Point (CEP) in the baryon chemical potential ($\mu_B$) and temperature ($T$) plane is a challenging goal from theoretical and experimental perspectives. 
It is well established that in some regions of density and temperature, the so-called chiral symmetry is restored from the hadronic matter to the quark-gluon plasma phase (QGP): LQCD with 2+1 light flavors shows a transition in the $T$-axis as a crossover with a pseudo-critical temperature $T_c\simeq 150-156$MeV~\cite{aoki2006qcd, PhysRevLett.113.082001,PhysRevD.85.054503} (see Ref.~\cite{guenther2022overview} for an overview of recent progress in the phase diagram).  Recent determination of the transition temperature with increased precision at finite baryon density can be found in Refs.~\cite{hotqcd2019a,hotqcd2019b}, with $T_c =156.5\pm1.5$ MeV at $\mu_B = 0$. Moreover, many theoretical and phenomenological models predict a first-order transition line in the low $T$ and large $\mu_B$ region~\cite{PhysRevD.41.1610,PhysRevD.49.426,PhysRevD.97.034015, PhysRevC.71.044904,ayala2016chiral,PhysRevD.71.114014}. Hence, the pass from a second order phase transition regime to one of first order defines the CEP location. }

Although the phenomenological description of the experimental data assumes conditions close to ideal thermalization, it is natural to assume that in a heavy-ion collision, there are stages where the thermal equilibrium cannot be demanded a priory~\cite{PhysRevC.83.034907,PhysRevC.81.064901,BLAIZOT1987847}. Indeed, the QGP-phase is preceded by the Glasma, which corresponds to a high gluon occupation system where its constituents are out of thermal equilibrium~\cite{LAPPI2006200,PhysRevD.95.094009,Venugopalan_2008}. Previous works have demonstrated that the signals of the glasma can be relevant to the final observables such as elliptic flow and photon invariant momentum distribution~\cite{ruggieri2013elliptic,PhysRevC.89.054914,PhysRevD.96.014023,ayala2020centrality}. Then, it is interesting to inquire if the non-equilibrium stages related to the QGP formation may modify the findings about the chiral symmetry restoration. 

As an attempt to describe situations outside of thermal equilibrium, Super statistics (SS) is one of the most attractive frameworks for describing the non-equilibrium dynamics of complex systems. Beck and Cohen introduced the SS as an extension of the equilibrium Boltzmann statistics by considering fluctuations in some extensive parameter $\tilde{\beta}$~\cite{beck2003superstatistics,beck2004superstatistics,beck2009recent}. Such a parameter can be identified as the inverse temperature, vorticity, friction constant, volatility in finance, or a quantity whose space-time fluctuations are much larger than the typical relaxation time of the local dynamics. Previous works have shown that a SS description of the temperature and baryon chemical potential impact the CEP location~\cite{ayala2018superstatistics,ayala2020fluctuating}.{ On the other hand, particular choices of SS are related to non-extensive thermodynamic scenarios~\cite{PhysRevE.104.024139,castano2022entropy}, so that SS may serve as a link between the thermal fluctuations~\cite{PhysRevD.93.014029,PhysRevD.96.034011} and the studies with non-extensive statistical mechanics~\cite{PhysRevD.101.096006,rozynek2016nonextensive,ishihara2016chiral} or explicit volume dependence~\cite{PhysRevD.104.074035,klein2017modeling,braun2012phase,PhysRevC.84.011903,Palhares_2011,Palhares_2010} for the QCD phase diagram.}

This work presents the interplay between the thermal fluctuations and the finite-volume effects in the CEP location in the L$\sigma$M framework. First, the out-of-equilibrium situation is modeled with the $\chi^2$ distribution function so that the fluctuations of the inverse temperature $\beta$ are encoded in the parameter $q$. The CEP's volume dependence arises from the definition of the Boltzmann partition function, which enters explicitly into the SS-thermodynamic potentials within a Tsallis-like prescription. The paper is organized as follows: In Sec.~\ref{sec:SS}, we present a summary of the SS and its relation with the non-extensive Tsallis thermodynamics. Next, in Sec.~\ref{sec:SSOmega}, the SS-effective potential density is found in its general form as a function of its equilibrium counterpart. Then, in Sec.~\ref{sec:LSM}, we present the effective density potential (in equilibrium) from the L$\sigma$M coupled to quarks. The results and discussion are presented in Sec.~\ref{sec:Results}. Finally we summarize and give an outlook of the analysis in Sec.~\ref{sec:Concl}.

\section{Super Statistics}\label{sec:SS}
The SS is based on the idea that some intensive parameter $\widetilde{\beta}$ may fluctuate by following a certain probability distribution function $f(\widetilde{\beta})$. By assuming that the system passes thought equilibrium states $e^{-\widetilde{\beta}\hat{H}}$, it is possible to construct a modified Boltzmann factor from the superposition of two statistics: one referring to the local equilibrium, and other due the fluctuations. Explicitly, this modified Boltzmann factor reads:
\bea
\hat{B}\equiv\int_0^\infty d\widetilde{\beta}f(\widetilde{\beta})e^{-\widetilde{\beta}\hat{H}},
\eea
where $\hat{H}$ is the Hamiltonian of the system. At this level, SS is an ansatz, and therefore, it cannot be taken as an first principles model for thermodynamic fluctuations.

Several models for $f(\widetilde{\beta})$ are used, depending of the physical situation to be modeled. For example, the most used are:
\begin{itemize}
    \item {\bf Uniform distribution}: It is the simplest distribution, given by
    \bea
    f(\tilde{\beta})=\frac{1}{b},
    \eea
    with $b$ a constant. 
    \item {\bf Multi-level distribution}: This distribution may appear when the system passes through several stages of Boltzmann-like equilibrium, each with equal probability. Its mathematical form is:
    \bea
    f(\tilde{\beta})=\frac{1}{N}\sum_{k=1}^N\delta(\tilde{\beta}-\beta_k).  
    \eea

    \item {\bf Gamma distribution}: The Gamma or $\chi^2$ distribution is given by
   \bea
   {f(\tilde{\beta})=\frac{1}{b\Gamma(c)}\left(\frac{\tilde{\beta}}{b}\right)^{c-1}e^{-\tilde{\beta}/b},}
   \label{chisquared}
   \eea
where $b$ and $c$ are free parameters.

\item {\bf Log-normal distribution}: This distribution function is given by
\bea
f(\tilde{\beta})=\frac{1}{\sqrt{2\pi}\tilde{\beta}u}\exp\left[-\frac{\log^2\left(\tilde{\beta}/v\right)}{2u^2}\right],
\eea
where $u$ and $v$ are free parameters.
    \item {\bf F-distribution}: For positive integers $v$, and $w$ and $b>0$, the F-distribution is defined as:
    \bea
    f(\tilde{\beta})=\frac{\Gamma[(v+w)/2]}{\Gamma(v/2)\Gamma(w/2)}\left(\frac{bv}{w}\right)^{v/2}\frac{\tilde{\beta}^{\frac{v}{2}-1}}{\left(1+\frac{bv}{w}\tilde{\beta}\right)^{(v+w)/2}}.\nn\\
    \eea
    
    The latter gives the Tsallis distribution in the $\tilde{\beta}$-space when $v\to2$.
\end{itemize}

In this work, in order to obtain analytical results for $\hat{B}$, we use the  gamma-distribution function. An interesting case is provided when its parameters are chosen as
\bea
{bc=\beta=1/(k_\text{B}T),~\text{and}~c=-1/(1-q),}
\eea
so that $\hat{B}$ can be written in terms of a $q$-exponential, namely:
\bea
\hat{B}(\beta)=e_q^{-\beta\hat{H}},
\label{Bq}
\eea
where the $q$-exponential is defined as
\bea
e_q^x\equiv[1+(1-q)x]^{1/(1-q)}.
\eea

From Eq.~(\ref{Bq}) it is possible to define a density operator as follows:
\bea
\hat{\rho}=\frac{1}{\mathcal{Z}}e_q^{-\beta\hat{H}},
\eea
where $\mathcal{Z}$ is the partition function given by
\bea
\mathcal{Z}=\text{Tr}\hat{\rho},
\label{rhoSS}
\eea
and from which the thermodynamic properties can be defined. In particular, in order to describe the QCD chiral symmetry restoration, we are interested in the L$\sigma$M effective potential density which is described in the next section.

\section{Super Statistical effective potential density}\label{sec:SSOmega}




As is pointed out in previous works, one can explore the similarities of Eq.~(\ref{Bq}) within the non-additive Tsallis thermodynamics~\cite{PhysRevE.104.024139,castano2022entropy}. The idea consists in preserving the Legendre structure (LS) of thermodynamics, which is lost in the Boltzmann-like formulation when the $q$-exponential is involved~\cite{tsallis1998role}. In the Tsallis prescription, the effective potential density is given by
\bea
\Omega_\text{T}=-\frac{1}{V\beta}\ln_q\mathcal{Z},
\label{OmegaT}
\eea
where the subscript ``T'' is for Tsallis, and $\ln_q x$ is the $q$-logarithm, defined as:
\bea
\ln_q x=\frac{x^{1-q}-1}{1-q}.
\eea

Preservation of LS is a desirable feature of a thermodynamic theory, because it is related to a increasing entropy and positive definite specific heat~\cite{scarfone2016consistency,plastino1997universality,castano2020super}.




\section{Super Statistical partition function}\label{sec:SSZ}
\begin{figure}
    \centering
    \includegraphics[scale=0.55]{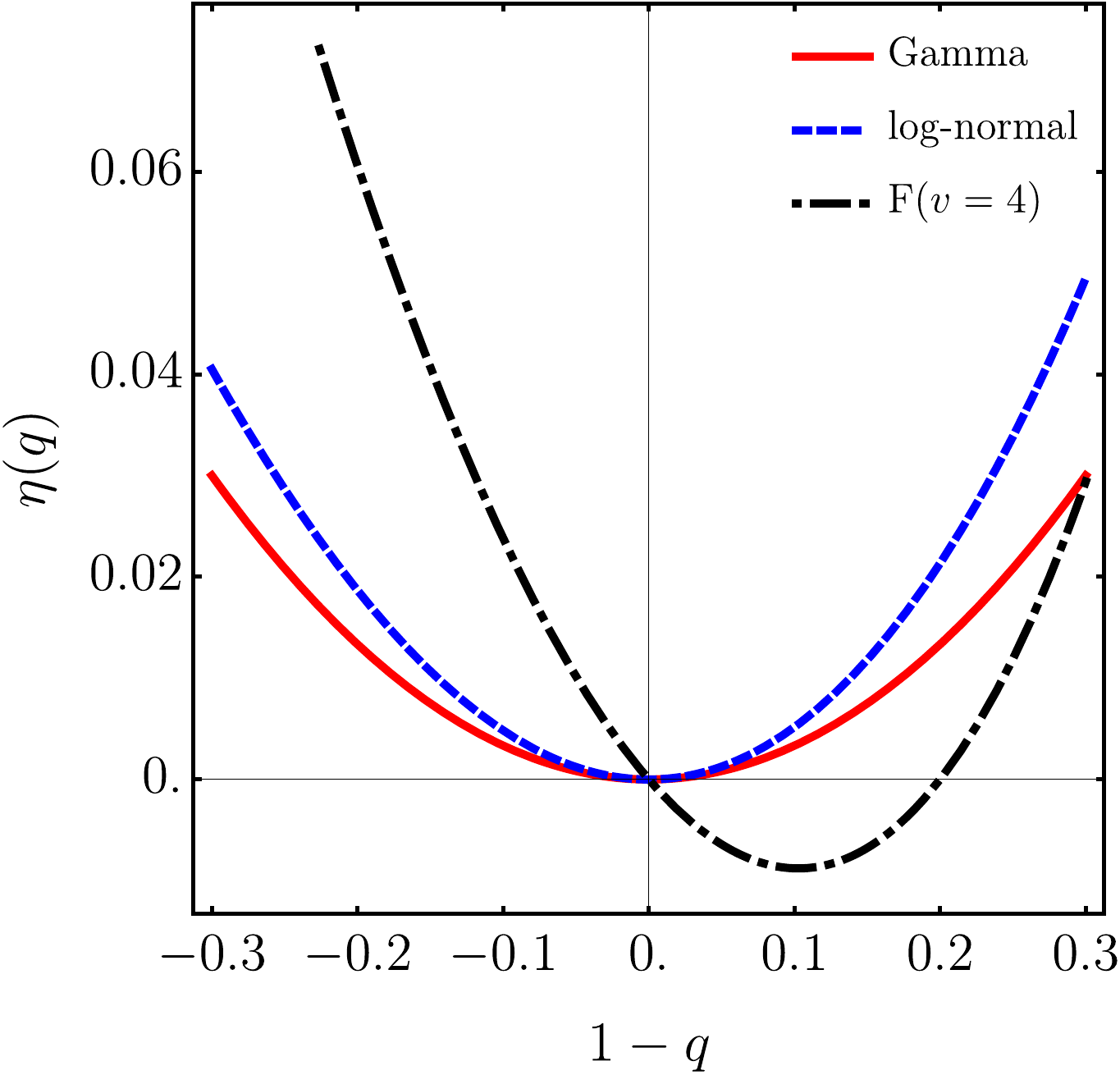}
    \caption{Function $\eta(q)$ for three distribution functions.}
    \label{fig:eta}
\end{figure}

{
In this section, we explore two approximations of the modified Boltzmann factors in order to obtain analytical results. The first one, done only for the Gamma distribution function, is provided by series expansion in powers of $(q-1)$ up to order $\mathcal{O}(q-1)^2$, which allows computing the corrections due to the finite size and the thermal fluctuation in the CEP's location. The second one is given in terms of an expansion in powers of $\beta\hat{H}$ from where the expressions for several distribution functions are different, providing
more information on the underlying complex dynamics for the fluctuations in each case.

At this point, it is necessary to comment on the validity of the expansion, which assumes small $\beta\hat{H}$. In fact, in the near to second-order phase transition, the fields are small given that the fermions approach the chiral limit where $gv<T$. This behavior is captured by the effective equilibrium potential of Sec.~\ref{sec:LSM} from which the SS correction are computed. On the other hand, near a first-order phase transition, there is a discontinuity in the order parameter $v$ and the fields
are not necessarily small, and an effective potential at low temperatures needs to be computed. Nevertheless, as is shown in Ref.~\cite{ayala2016chiral}, the results obtained from approximations of high and low temperatures are closely related by a change in the coupling constants of fermions and bosons. Then, as a first approximation, the product $\beta\hat{H}$ might be considered small. Corrections due to the low-temperature regime are in process, and we will report them elsewhere.
\subsection{Expansion in powers of $q-1$}\label{subsec:Expansioninq}

In order to find a expression for the SS-partition function, we expand Eq.~(\ref{rhoSS}) around $q=1$, so that up to order $\mathcal{O}(q^2)$:
\bea
\mathcal{Z}&\approx&\Z_0+\frac{q-1}{2}\beta^2\frac{\partial^2\Z_0}{\partial\beta^2}\nn\\
&+&\frac{(q-1)^2}{24}\left(8\beta^3\frac{\partial^3\Z_0}{\partial\beta^3} +3\beta^4\frac{\partial^4\Z_0}{\partial\beta^4} \right),
\label{Zq}
\eea
where $\Z_0$ is the Boltzmann partition function given by:
\bea
\mathcal{Z}_0=\exp\left(-V\beta\Omega_0\right),
\label{Z0}
\eea
with $\Omega_0$ is the equilibrium effective potential density, and $V$ is the volume of the system. Note that the derivatives of $\Z_0$ introduce a non-trivial volume dependence, namely,
%

\begin{widetext}
\bea
\Z&\approx&e^{-\beta V\Omega_0}\left\{ 1+\frac{q-1}{2} \beta^2 \left[ V^2 \left(  \Omega_0 +\beta  \frac{\partial \Omega_0}{\partial \beta} \right)^2 -2V\frac{\partial \Omega_0}{\partial \beta} -\beta V \frac{\partial^2 \Omega_0}{\partial \beta^2} \right] \right.   \nn\\
&+&\left.\frac{(q-1)^2}{3}\beta^3 V \left[ -V^2 \left( \Omega_0 +\beta\frac{\partial\Omega_0 }{\partial \beta}  \right)^3-3\frac{\partial^2\Omega_0 }{\partial \beta^2} +3V\left(\Omega_0 +\beta \frac{\partial\Omega_0 }{\partial \beta} \right)\left(2\frac{\partial\Omega_0 }{\partial \beta}+\beta \frac{\partial^2 \Omega_0 }{\partial \beta^2} \right)-\beta\frac{\partial^3 \Omega_0 }{\partial \beta^3} \right]   \right.   \nn\\
&+&\left.\frac{(q-1)^2}{8} \beta^4 \left[  V^4 \left( \Omega_0 +\beta \frac{\partial\Omega_0 }{\partial \beta} \right)^4 -6V^3 \left(\Omega_0 +\beta \frac{\partial\Omega_0 }{\partial \beta} \right)^2 \left( 2\frac{\partial\Omega_0 }{\partial \beta} +\beta \frac{\partial^2 \Omega_0 }{\partial \beta^2} \right)+3V^2 \left( 2 \frac{\partial\Omega_0 }{\partial \beta} +\beta \frac{\partial^2 \Omega_0 }{\partial \beta^2} \right)^2 \right. \right.   \nn\\
&-&\left.\left. 4V \frac{\partial^3\Omega_0 }{\partial \beta^3} +4V^2 \left( \Omega_0+\beta \frac{\partial\Omega_0 }{\partial \beta} \right)\left(3 \frac{\partial^2\Omega_0 }{\partial \beta^2}+\beta \frac{\partial^3 \Omega_0 }{\partial \beta^3} \right) -\beta V \frac{\partial^4 \Omega_0 }{\partial \beta^4} \right]     \right\}.
\eea
\end{widetext}

The latter partition function has two interesting features: it includes thermal fluctuations within the $q$-parameter, and links the non-extensive scenario with an explicit volume dependence. 

\subsection{Differences with other distribution functions from an expansion in powers of $\beta\hat{H}$}
As was commented in Eq.~(\ref{ExpansionInSigma}), the correction up to order $\mathcal{O}(\sigma\hat{H})^2$ is the same for most of the distribution functions when a series expansion is done in powers of $\sigma^2$ or $q-1$. However, there is another way to expand the modified Boltzmann factor, which corresponds to powers of $\beta\hat{H}$ introducing noticeable differences between the expressions. In general, such a expansion can be written as~\cite{beck2003superstatistics}:
\bea
\hat{B}(\hat{H})=e^{-\beta\hat{H}}\left[1+\frac{q-1}{2}\beta^2\hat{H}^2+\eta(q)\beta^3\hat{H}^3+\ldots\right],\nn\\
\eea
where:
\bea
\eta(q)= \begin{cases}
    0, & \text{for uniform and 2-level dist.}  \\
    -\frac{1}{3}(q-1)^2, &\text{for Gamma dist.}\\
    -\frac{1}{6}\left(q^3-3q+2\right), & \text{for log-normal dist.}\\
    -\frac{1}{3}\frac{(q-1)(5q-6)}{3-q}, & \text{for } F\text{ dist. with }v=4 .
  \end{cases},
\eea
and the $q$-index is given for each distribution by:
\bea
q=\begin{cases}
w, & \text{for log-normal dist.}\\
q=1+\frac{2(v+w-2)}{v(w-4)}, & \text{for } F\text{ distribution}  \\
\end{cases}
\eea
Figure~\ref{fig:eta} shows the differences of $\eta(q)$ for each distribution function. 

Note that with this power series we ignore in Eq.~(\ref{Zq}) the term
\bea
\frac{(q-1)^2}{24}\left(3\beta^4\frac{\partial^4\Z_0}{\partial\beta^4}\right),
\eea
which is proportional to $\beta^4\hat{H}^4$. However, although that term seems essential in the analytical expression of the potential, the numerical calculations show that in the present work, it is negligible when compared with
\bea
\frac{(q-1)^2}{24}\left(8\beta^3\frac{\partial^3\Z_0}{\partial\beta^3}\right).
\eea

Then, the results for the Gamma distributions are the same that the related in Sec.~\ref{subsec:Expansioninq}.

From the above, the general form of the partition function will be:
\begin{widetext}
\bea
\Z&\approx&e^{-\beta V\Omega_0}\left\{ 1+\frac{q-1}{2} \beta^2 \left[ V^2 \left( \Omega_0 +\beta \frac{\partial \Omega_0}{\partial \beta} \right)^2 -2V\frac{\partial \Omega_0}{\partial \beta} -\beta V \frac{\partial^2 \Omega_0}{\partial \beta^2} \right] \right.   \nn\\
&-&\left.\eta(q)\beta^3 V \left[ -V^2 \left( \Omega_0 +\beta\frac{\partial\Omega_0 }{\partial \beta}  \right)^3-3\frac{\partial^2\Omega_0 }{\partial \beta^2} +3V\left(\Omega_0 +\beta \frac{\partial\Omega_0 }{\partial \beta} \right)\left(2\frac{\partial\Omega_0 }{\partial \beta}+\beta \frac{\partial^2 \Omega_0 }{\partial \beta^2} \right)-\beta\frac{\partial^3 \Omega_0 }{\partial \beta^3} \right] \right\}.
\label{Zdifdist}
\eea
\end{widetext}
}

\section{Equilibrium effective potential of linear sigma model coupled to quarks}\label{sec:LSM}
\begin{figure*}
    \centering
    \includegraphics[scale=0.55]{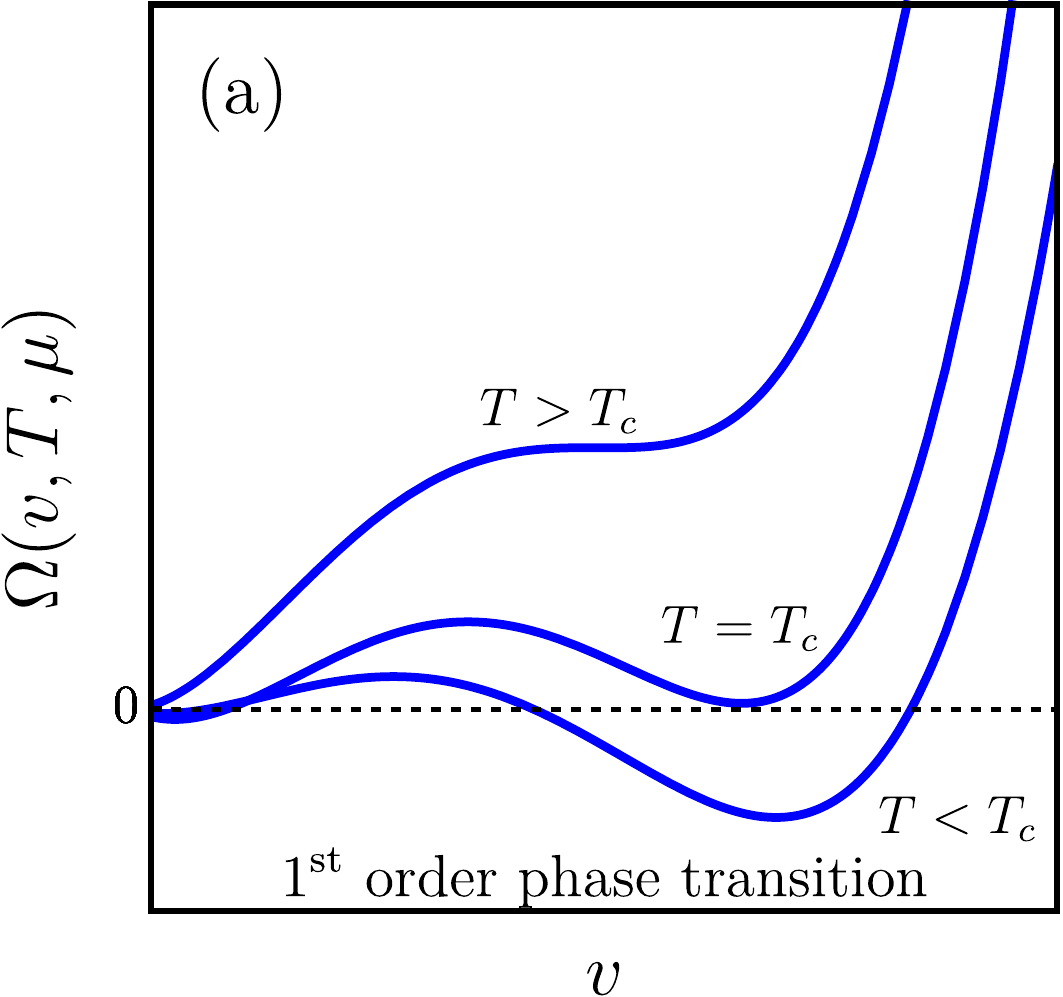}
    \hspace{0.8cm}
    \includegraphics[scale=0.55]{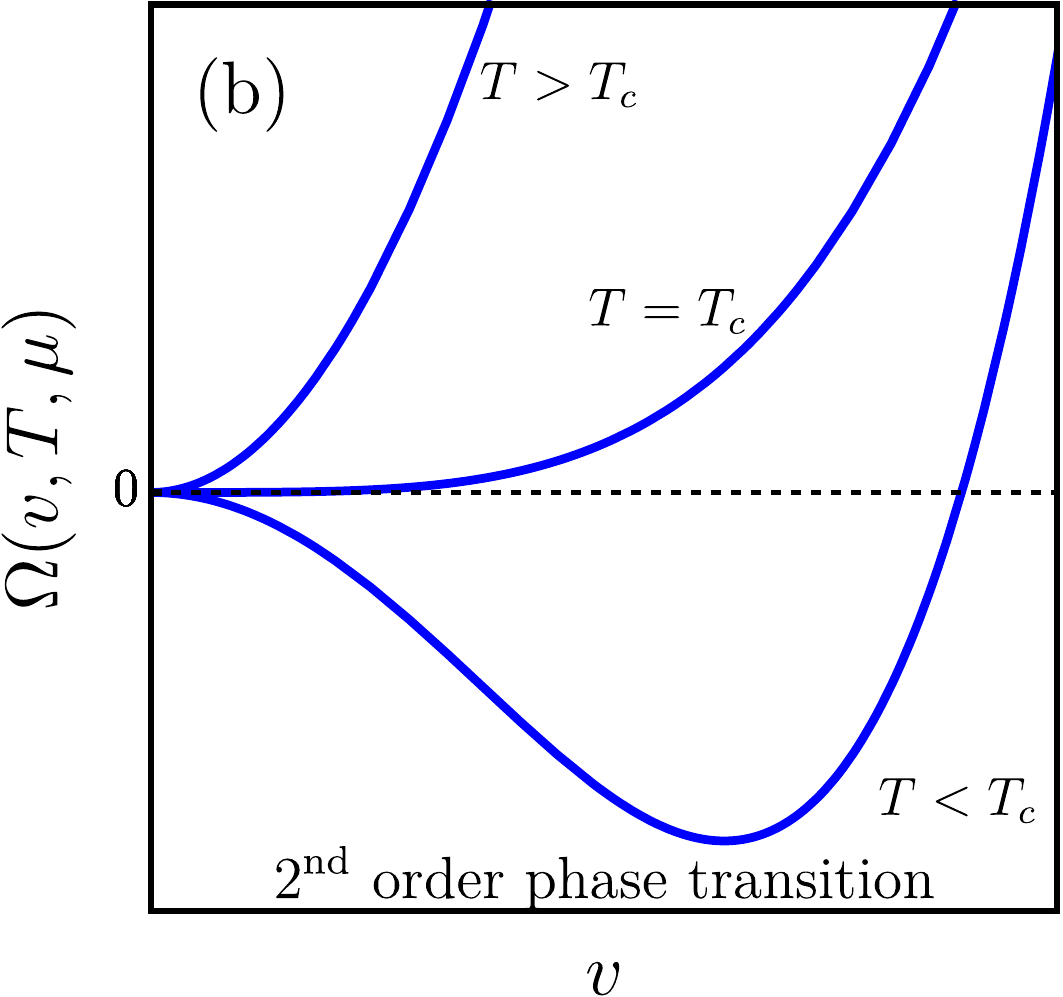}
    \caption{Shapes of the effective potential as function of the order parameter $v$ and for temperatures lowers, equal, and higher than the transition temperature $T_c$. }
    \label{fig:transitionorder}
\end{figure*}
In Sec.~\ref{sec:SSZ}, the SS effective potential density is written in terms of the equilibrium potential $\Omega_0$. To obtain it, and in order to compute some features of the QCD phase diagram, we use the LSMq whose Lagrangian density is given by~\cite{ayala2016chiral}:
\bea
\mathcal{L}&=&\frac{1}{2}(\partial_\mu\sigma)^2+\frac{1}{2}(\partial_\mu\boldsymbol{\pi})^2+\frac{a^2}{2}(\sigma^2+\boldsymbol{\pi}^2)-\frac{\lambda}{4}(\sigma^2+\boldsymbol{\pi}^2)^2\nn\\
&+&\ii\bar{\psi}\gamma^\mu\partial_\mu\psi-g\bar{\psi}(\sigma+\ii\gamma_5\boldsymbol{\tau}\cdot\boldsymbol{\pi})\psi,
\eea
where the mass parameter $a^2$, and the couplings $\lambda$ and $g$ are positive constants, $\psi$ is an SU(2) isospin doublet, and $\boldsymbol{\pi}$ and $\sigma$ are an isospin triplet and singlet, respectively. We take the neutral pion $\pi^0$ as the third component of the triplet, and the charged pions as
\bea
\pi_\pm=\frac{1}{2}(\pi_1\mp\ii\pi_2).
\eea

It is well established that this model admits spontaneous symmetry breaking, which can be realized over the $\sigma$ field when it develops a vacuum expectation value $v$. This mechanism is obtained from the shift:
\bea
\sigma\to\sigma+v,
\eea
so that
\bea
\mathcal{L}&=&\frac{a^{2}}{2} v^{2}-\frac{\lambda}{4} v^{4}\nn\\
&-&\frac{1}{2}\left(\partial_{\mu} \sigma\right)^2-\frac{1}{2}\left(3 \lambda v^{2}-a^{2}\right) \sigma^{2}\nn\\
&-&\frac{1}{2}\left(\partial_{\mu}\boldsymbol{\pi}\right)^2-\frac{1}{2}\left(\lambda v^{2}-a^{2}\right) \boldsymbol{\pi}^{2}\nn\\
&+&\ii \bar{\psi} \gamma^{\mu} \partial_{\mu} \psi-g v \bar{\psi} \psi+\mathcal{L}_{I}^{b}+\mathcal{L}_{I}^{f}
\label{eq:LagrangianLsM}
\eea
where
\begin{subequations}
\bea
\mathcal{L}_{I}^{b}=-\frac{\lambda}{4}(\sigma^2+\boldsymbol{\pi}^2)^2,
\eea
and
\bea
\mathcal{L}_{I}^{f} =-g \bar{\psi}\left(\sigma+\ii \gamma_{5} \boldsymbol{\tau} \cdot \boldsymbol{\pi}\right) \psi.
\eea
\end{subequations}

The latter equations implies that after symmetry breaking, the involved fields acquire mass given by
\begin{subequations}
\bea
m_\sigma^2=3\lambda v^2-a^2,
\eea
\bea
m_\pi^2=\lambda v^2-a^2,
\eea
\bea
{m_f=gv.}
\label{quarkmass}
\eea
\label{FieldMasses}
\end{subequations}

To compute the equilibrium effects of the finite temperature and density in the chiral symmetry restoration, we use the effective potential and the self-energy up to the ring diagrams, so that in a high-temperature limit where the quark masses are small, they are~\cite{AYALA201577,ayala2016chiral}:
\begin{widetext}
\bea
\Omega_0&=&-\frac{a^2}{2}v^2+\frac{\lambda}{4}v^4+\sum_{i=\sigma,\boldsymbol{\pi}}\Bigg\{\frac{m_i^4}{64\pi^2}\left[\ln\left(\frac{16\pi^2T^2}{2a^2}\right)-2\gamma_\text{e}+1\right]-\frac{\pi^2T^4}{90}+\frac{m_i^2T^2}{24}-\frac{T}{12\pi}\left[m_i^2+\Pi(T,\mu)\right]^{3/2}\Bigg\}\nn\\
&-&\frac{N_c}{16\pi^2}\sum_{f=u,d}\Bigg\{m_f^4\Bigg[\ln\left(\frac{8\pi^2T^2}{a^2}\right)+\psi_0\left(\frac{1}{2}+\frac{\ii\mu}{2\pi T}\right)+\psi_0\left(\frac{1}{2}-\frac{\ii\mu}{2\pi T}\right)+1\Bigg]+8m_f^2T^2\left[\text{Li}_2\left(-e^{\mu/T}\right)\right.\nn\\
&+&\left.\text{Li}_2\left(-e^{-\mu/T}\right)\right]-32T^4\left[\text{Li}_4\left(-e^{\mu/T}\right)+\text{Li}_4\left(-e^{-\mu/T}\right)\right]\Bigg\},
\label{Omega0}
\eea
\end{widetext}
and
\bea
\Pi(T,\mu)&=&\frac{\lambda T^2}{2}\\
&-&\frac{N_fN_cg^2T^2}{\pi^2}\left[\text{Li}_2\left(-e^{\mu/T}\right)+\text{Li}_2\left(-e^{-\mu/T}\right)\right],\nn
\eea
where $\gamma_\text{e}$ is the Euler-Mascheroni constant, $\psi_0(x)$ is the digamma function, and $\text{Li}_n(x)$ is the poly-logarithm function of order $n$. Moreover, we take $N_c=3$ and $N_f=2$, as the number of colors and flavors, respectively. The coupling constants are fixed with conditions at $\mu=0$, where the physical boson masses (corrected by the self-energy) are:
\begin{subequations}
\bea
m_\sigma^2=3\lambda v^2-a^2+\frac{\lambda T^2}{2}+\frac{N_fN_cg^2T^2}{6},
\eea
and
\bea
{m_\pi^2=\lambda v^2-a^2+\frac{\lambda T^2}{2}+\frac{N_fN_cg^2T^2}{6},}
\eea
\label{SistemMasses}
\end{subequations}
so that in the transition temperature $T_c$, the masses vanishes. Then, by solving for $a$ from Eqs.~(\ref{SistemMasses}):
\bea
a=T_c\sqrt{\frac{\lambda}{2}+\frac{N_fN_cg^2}{6}}.
\eea

Now, from Eqs.~(\ref{FieldMasses}) is easy to check that:
\bea
a=\sqrt{\frac{m_\sigma^2-3m_\pi^2}{2}}
\eea

The present work takes into account only two quark flavors in the chiral limit, which allows
to compare with lattice simulations for $N_f = 2 + 1$, where the critical temperature at $\mu=0$ is $T\simeq170$ MeV~\cite{Maezawa_2007}. Then, the constants $\lambda$ and $g$ can be identified, and throughout the paper, we take them as $\lambda=0.86$, and $g=1.1$.

Our goal is to describe the chiral symmetry restoration where the quark mass vanishes so that the sigma's field vacuum expectation value $v$ is promoted as the order parameter for the transition. In that spirit, the phase transition and its order are identified from the effective potential shape in such a way that the order parameter has a discontinuity at the critical temperature $T_c$ for a first order phase transition. In contrast, in a second-order phase transition, it evolves continuously. Figure~\ref{fig:transitionorder} illustrates the potential shape for both situations. 

\begin{figure*}
    \centering
    \includegraphics[scale=0.65]{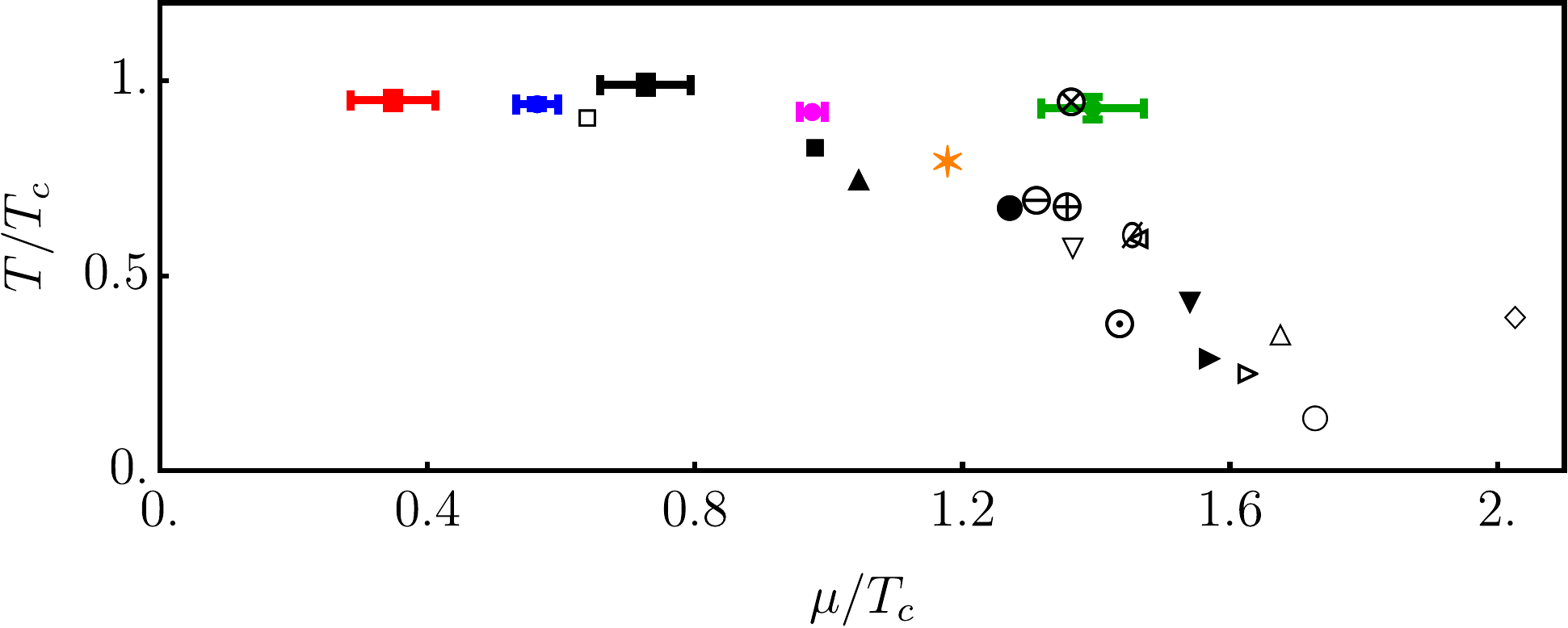}
    \caption{CEP's location from different approaches. The points with error bars are LQCD extensions from Refs.~\cite{PhysRevD.71.114014} (red),~\cite{PhysRevD.78.114503} (blue),~\cite{fodor2004critical} (black),~\cite{deForcrand:2006ec} (magenta), and~\cite{fodor2002lattice} (green). The black filled symbols correspond to the low-temperature approach of Eq.~\eqref{eq:LagrangianLsM}~\cite{ayala2016chiral,castano2021effects} when the coupling constants $(\lambda,g)$ are: $(2.2,1.7)$ for $\blacksquare$, $(2.4,1.65)$ for $\blacktriangle$, $(2.1,1.725)$ for $\bullet$, $(2.5,1.7)$ for $\blacktriangledown$, and $(2.7,1.6)$ for $\blacktriangleright$. The empty symbols are model predictions: $\ominus$ from Ref.~\cite{GAO2021136584}, $\varnothing$ from Ref.~\cite{PhysRevD.102.034027}, $\oplus$ from Ref.~\cite{PhysRevD.101.054032},  $\square$ from Ref.~\cite{ayala2018superstatistics}, $\bigotimes$ from Ref.~\cite{PhysRevD.88.056004},  $\bigtriangledown$ from Ref.~\cite{PhysRevD.77.014006}, $\bigodot$ from Ref.~\cite{PhysRevD.81.016007}, $\vartriangleleft$ from Ref.~\cite{PhysRevD.79.014018}, $\vartriangleright$ from Ref.~\cite{PhysRevD.78.034034}, $\bigtriangleup$ from Ref.~\cite{PhysRevD.77.096001}, $\bigcirc$ from Ref.~\cite{Ayala:2019skg}, and $\lozenge$ from Ref.~\cite{PhysRevD.77.065016}. The orange six-point star corresponds to the CEP located with Eq.~(\ref{Omega0}). (Figure adapted from Ref.~\cite{castano2021effects}).}
    \label{fig:CEPScomparison}
\end{figure*}

It is worth saying that the potential of Eq.~(\ref{Omega0}) is obtained by assuming that temperature is the dominant energy scale, i.e., $T>\mu>m_f$ or $T>gv$, so that the fermions are close to the chiral limit. Therefore, in a strict sense, the potential is not valid in regions where $\mu\gtrsim T$ where the first-order transition lines begin. {Nevertheless, Fig.~\ref{fig:CEPScomparison} shows the CEP's location computed from Eq.~\eqref{eq:LagrangianLsM} in a low-temperature approximation~\cite{ayala2016chiral,castano2021effects} (filled black symbols), LQCD calculations (poinst with error bars), other phenomenological models (white-empty symbols), and the computed from Eq.~(\ref{Omega0}) without super-statistics (six-point orange star). As can be noticed, working with Eq.~(29) provides good results according to the current state-of-the-art of the CEP location. Therefore, if the temperatures are not low-enough (we are not interested on the first-order transition lines but in the CEP's coordinates), our results can be regarded as a valid approximation.}

\section{Results and discussion}\label{sec:Results}

In the previous sections, we developed a formalism to describe the QCD chiral symmetry restoration within the L$\sigma$M coupled to quarks when the system has fluctuations in its temperature. Moreover, given the connection between SS and the Tsallis non-additive thermodynamics, our formalism links the out-of-equilibrium situation with an explicit volume dependence. This section explores the impact of thermal fluctuations and the system's size in the CEP location and the transition lines.

{
\subsection{Expansion in powers of $q-1$}
}
Before presenting our results, let us comment that the parameters $q$ and $V$ are independent. This is an important feature because the SS formalism can be interpreted as if the system is divided into $N$ subsystems, each with local equilibrium, and the distribution function $f(\widetilde{\beta})$ accounts for the spatial temperature distribution among them~\cite{beck2009recent,ayala2018superstatistics,ayala2020fluctuating}. The latter implies that the $q$-parameter can be identified as:
\bea
q=1+2/N.
\label{qandN}
\eea

Nevertheless, there is another interpretation of the SS: the system is not divided into subsystems, but it as a whole passes through several equilibrium stages. Then, the total volume remains unaltered, and the fluctuations acquire a temporal nature. We adopt this point of view.
\begin{figure}[h!]
    \centering
    \includegraphics[scale=0.55]{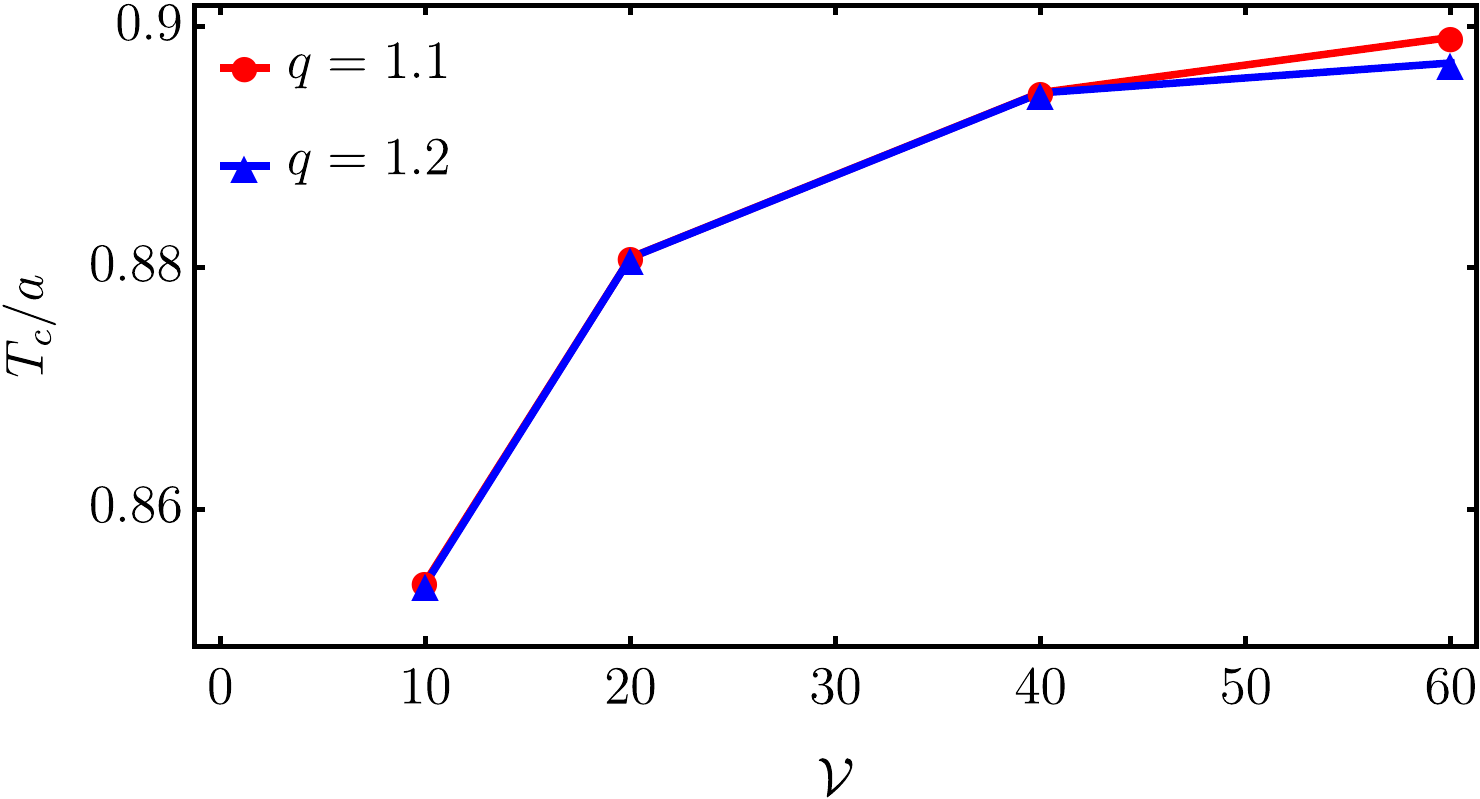}
    \caption{Pseudo-critial temperature $T_c$ at $\mu=0$, as a function of the dimensionless volume $\mathcal{V}=a^3V$ for $q=1.1$ and $q=1.2$. The reference temperature $T_c^0\approx0.9$ is the transition temperature at $\mu=0$ in the equilibrium case $q\to1$. The curves are obtained with the Tsallis prescription. A similar behavior is found with $q=0.8$, and $q=0.9$.}
    \label{fig:TcvsV}
\end{figure}

On the other hand, given that the SS is an ansatz, the $q$-parameter can be interpreted in several ways beyond the relation of Eq.~(\ref{qandN}). In fact, this parameter can be related to the variance of the inverse temperature fluctuations ~\cite{beck2003superstatistics}:
\bea
(q-1)\beta^2=\sigma^2\text{ or } q=\frac{\langle\tilde{\beta}^2\rangle}{\langle\tilde{\beta}\rangle^2},
\label{qandsigma}
\eea
so that it is possible to find a general expression for the modified Boltzmann factor given by:
\bea
\hat{B}=e^{-\beta\hat{H}}\left[1+\frac{1}{2}\sigma^2\hat{H}^2+\mathcal{O}\left(\sigma^3\hat{H}^3\right)\right],
\label{ExpansionInSigma}
\eea
where the variances (and therefore, the $q$-parameter) have different forms. For example, for the uniform, log-normal, and the F-distribution, the variances are respectively:
\begin{subequations}
\bea
\sigma^2=b^2/12,
\eea
\bea
\sigma^2=v^2e^{s^2}\left(e^{s^2}-1\right),
\eea
and
\bea
\sigma^2=\frac{2w^2(v+w-2)}{b^2v(w-2)^2(w-4)}.
\eea
\end{subequations}

{Note that Eq.~(\ref{qandsigma}) defines only positive values of $q-1$, and therefore, $q\geq1$. Nevertheless, we aim to connect with a non-extensive scenario, i.e., the Tsallis statistics in which a priory, there are no restrictions over $q$. Hence, in order to present general results (valid for both points of view), we explore the range $0.8\leq q\leq 1.2$ (values outside of this interval imply more terms in the $q-1$ series expansion).}

\begin{figure}[h!]
    \centering
     \includegraphics[scale=0.55]{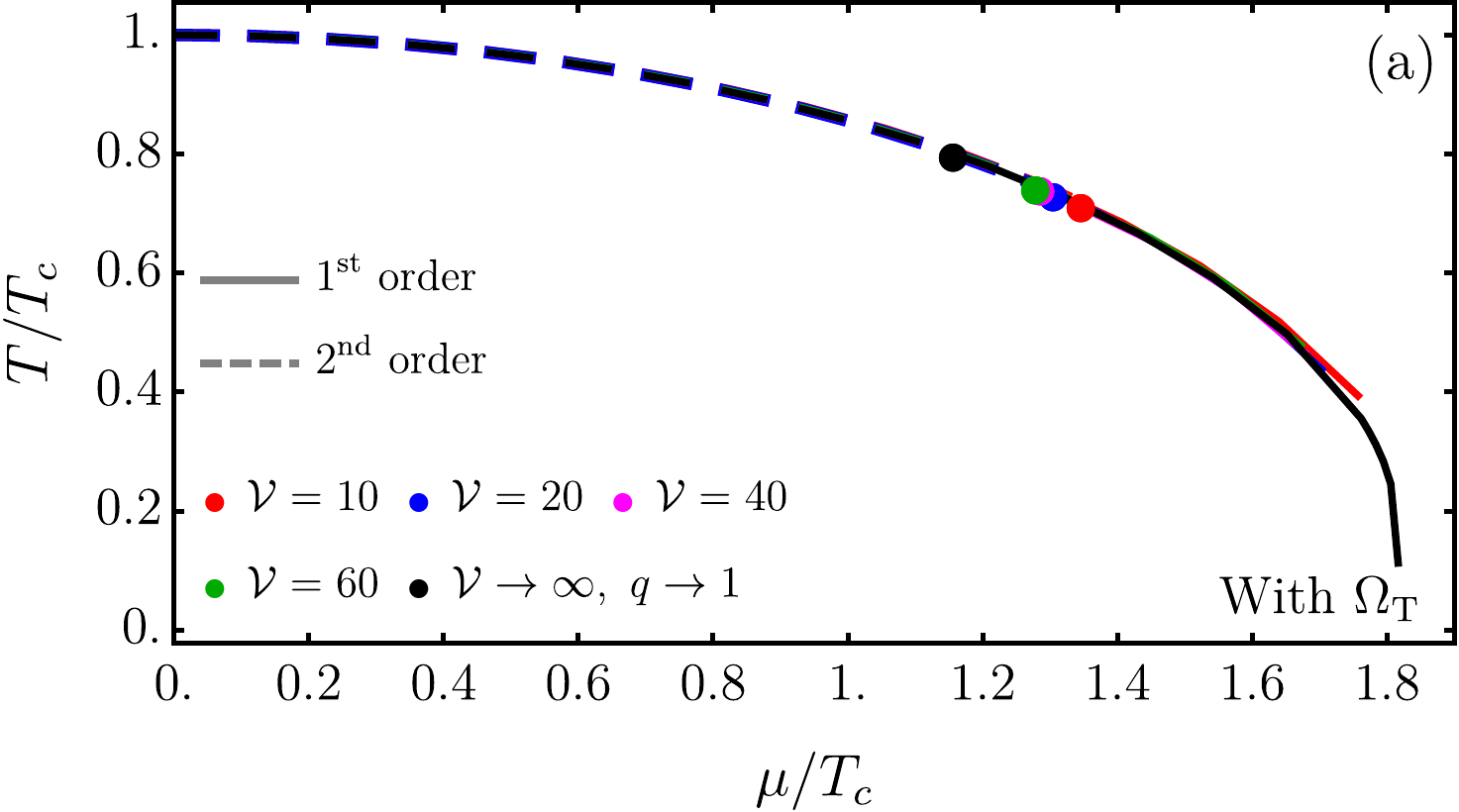}\\
    \vspace{0.3cm}
    \includegraphics[scale=0.55]{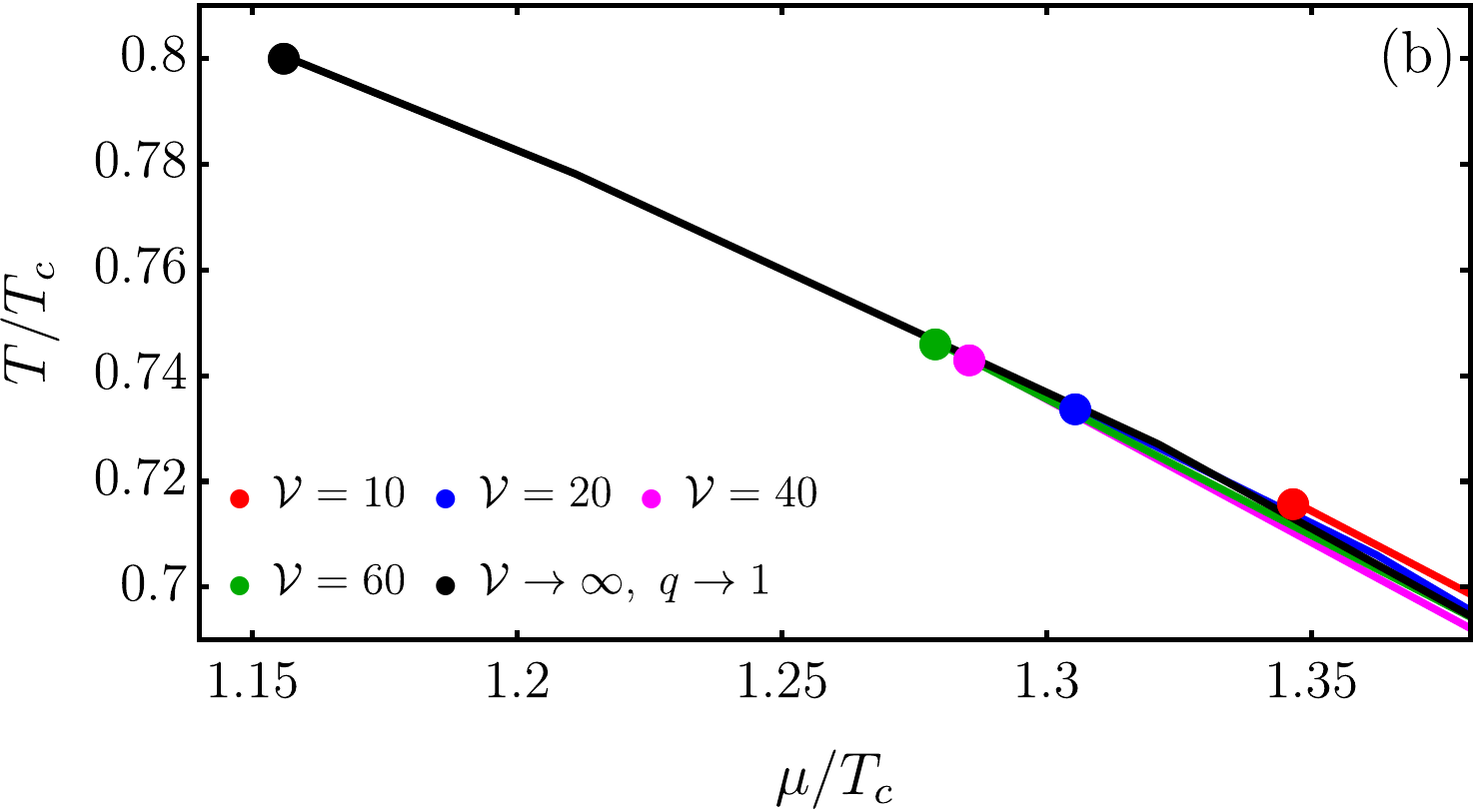}
    \caption{Effective QCD phase diagram obtained from the potential $\Omega_\text{T}$ for $q=1.2$ and several values of $\mathcal{V}$. The points are the CEP's location for each volume. The values of $T$ and $\mu$ in the critical line are normalized to their own $T_c$ which is volume-dependent. Panel (a) is the full phase diagram, and panel (b) is a zoom in order to appreciate the volume effects in the CEP (only the first-order critical lines are shown.).}
    \label{fig:PhaseDiagramTsallis}
\end{figure}

Figure~\ref{fig:TcvsV} shows the pseudo-critical temperature $T_c$ (at $\mu_B=0$) as a function of the thermal fluctuation's parameter $q$, and the dimensionless volume $\mathcal{V}\equiv a^3V$. In order to visualize the effects of the variables, we scaled $T_c$ with the pseudo-critical temperature at thermal equilibrium $T_c^0$. As can be noticed, the fluctuations in temperature reproduce almost the same results for $q=1.1$ and $q=1.2$ (the same behavior is found with $q=0.8$, and $q=0.9$). Still, for small system size, the transition temperature decreases around 20$\%$ from the value found in the equilibrium situation. Note that a similar scaling with the volume is found in Ref.~\cite{PhysRevD.104.074035} for a finite box model by using lattice Yang–Mills theory and Dyson–Schwinger equations for 2+1 quark flavors. 

Figure~\ref{fig:PhaseDiagramTsallis} presents the effective QCD phase diagram in the plane $\mu-T$ (recall $\mu_B=3\mu$) computed from Eq.~(\ref{OmegaT}) for several values of $\mathcal{V}$ with $q=1.2$. The dashed curves represent the crossover or second-order phase transition lines, whereas the continuous are the first-order critical regions. The dots are the CEP location when the critical values of $T$ and $\mu$ are scaled with the respective $T_c$ for each $\mathcal{V}$. The finite volume effect is translated to move the the CEP toward smaller temperatures and larger chemical potentials, which correspond to a change about $10\%$-$17\%$ for $\mu$, and 10$\%$-12$\%$ for $T$. Again, the findings are in agreement with the findings of with Ref.~\cite{PhysRevD.104.074035}. Moreover, the results are quantitatively the same when variations in $q$ are implemented, which indicates that the transition lines and the CEP location are robust against thermal fluctuations but have a considerable dependence on the size of the system.

{
\subsection{Expansion in powers of $\beta\hat{H}$}

\begin{figure}
    \centering
    \includegraphics[scale=0.71]{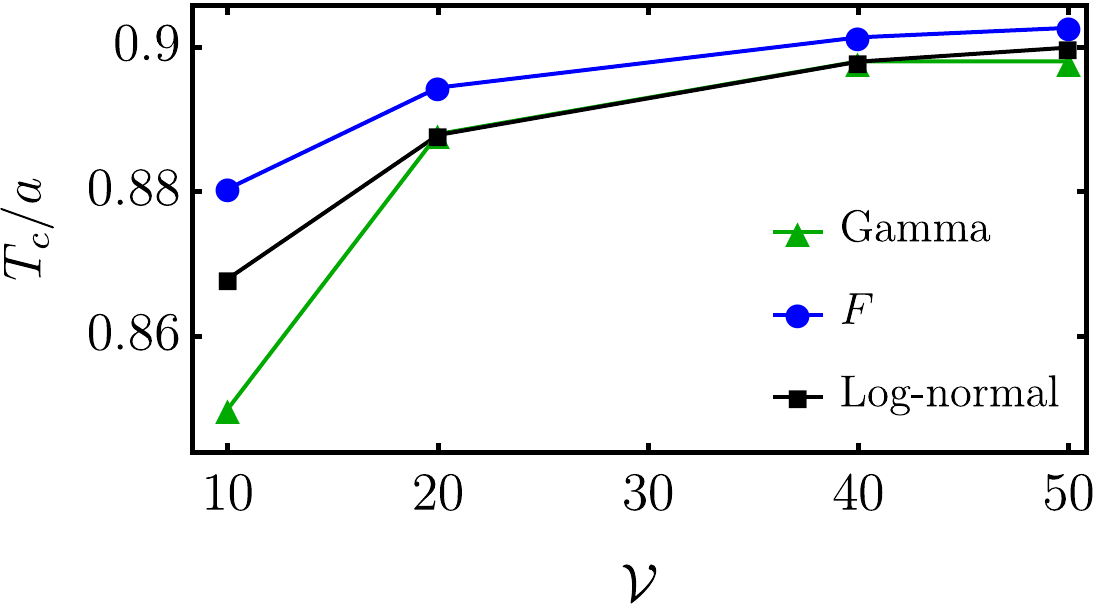}
    \caption{Pseudo-critial temperature $T_c$ at $\mu=0$, as a function of the dimensionless volume $\mathcal{V}=a^3V$ for $q=1.2$ and different distribution functions.}
    \label{fig:TcvsV2}
\end{figure}

\begin{figure}
    \centering
    \includegraphics[scale=0.68]{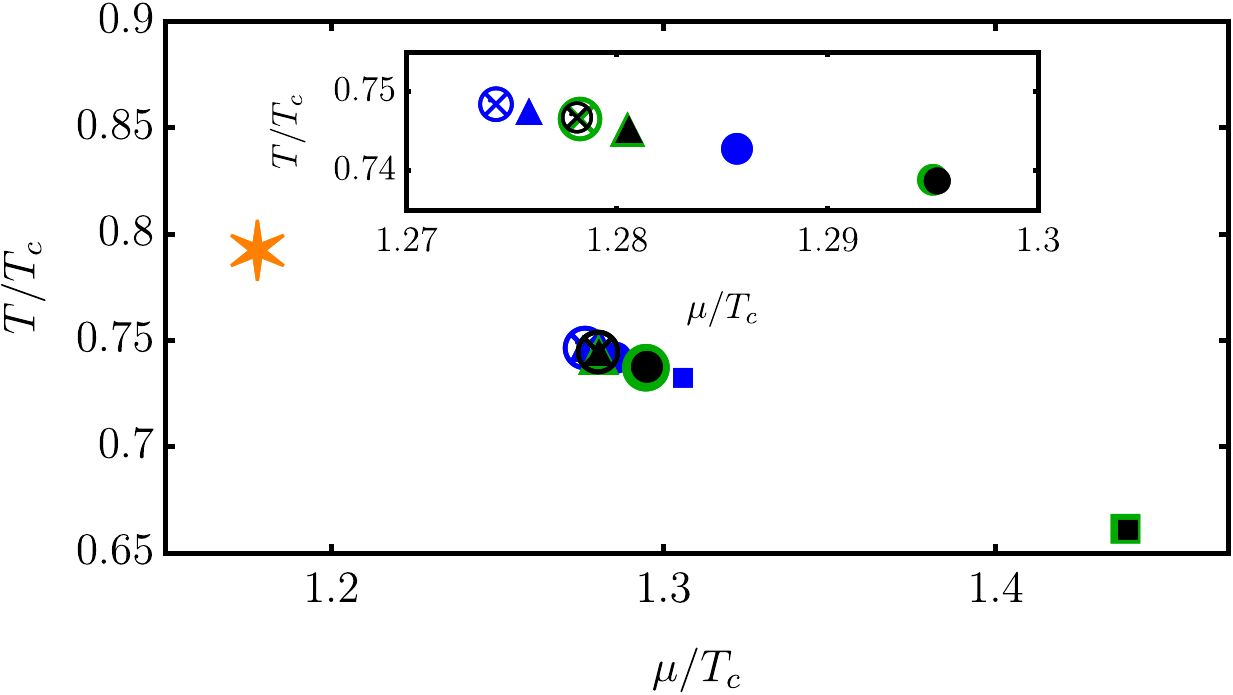}
    \caption{CEP's locations obtained from Eq.~(\ref{Zdifdist}) for different distribution functions: Gamma (green), $F$ (blue), and Log-normal (black). The symbols represent the dimensionless volume: $\otimes$ for $\mathcal{V}=50$, $\blacktriangle$ for $\mathcal{V}=40$, $\bullet$ for $\mathcal{V}=20$, and $\blacksquare$ for $\mathcal{V}=10$. The yellow star is the pure Boltzmann result. The inset shows a zoom in order to appreciate the CEP deviations of each distribution function. The results are independent of $q$, and here we present $q=1.2$.}
    \label{fig:CEPdiferentesdist}
\end{figure}

Finally, in order to compare the differences between super-statistics prescriptions, Figs.~\ref{fig:TcvsV2} and~\ref{fig:CEPdiferentesdist} show the changes of $T_c$ with $\mathcal{V}$, and the CEP's locations obtained from Eq.~(\ref{Zdifdist}), both for the Gamma, $F$, and Log-normal distribution functions. Once again, the results seem independent of the value of $q$, which indicates that in our series expansions, the temperature fluctuations are irrelevant, and the chiral symmetry restoration is robust in the present approximation. However, the $T_c$ and CEP are sensitive to the system's volume but the deviations from one distribution function to another are minor. In fact, only the $F$ distribution has noticeable differences when compared with the Gamma and the Log-normal results, which is similar to the observed in Figure \ref{fig:eta}. This behavior is not surprising: the series expansion of Eq.~(\ref{Zdifdist}) is made in powers of $\beta\hat{H}$, which is a tiny quantity near the transition. Then, in the critical lines, one may expect similar results.
}

\section{Summary and Conclusions}\label{sec:Concl}

In this work, we have used the L$\sigma$M coupled with quarks to locate the CEP in the effective QCD phase diagram by considering the finite-size effects produced by thermal fluctuations. The SS provides the connection between the out-of-equilibrium condition and the non-extensive thermodynamics with the $\chi^2$-distribution function so that its free parameters resemble the Tsallis statistics. Therefore, we end with an effective thermodynamic potential as a function of the thermal-fluctuations parameter $q$ and explicit volume dependence. To appreciate the finite size effects, we adopted the SS interpretation in which the distribution function models the system as a whole, passing for several stages close to the thermal equilibrium. Moreover,  given the universality of the modified Boltzmann factor, we expanded the range o values of $q$ by demanding its nature as a parameter into an ansatz. With this, several distribution functions can be considered.  

We find that the pseudo-critical temperature $T_c$ at $\mu=0$ is considerably changed when the system's volume decreases (around a $7\%$ for the smaller volume considered). Nevertheless, the temperature fluctuations do not represent significant changes in the value of $T_c$. The same situation is found in the CEP location: it moves towards high values of $\mu$ and lower temperatures, so that the former may change in a range of the $10\%-17\%$, and the latter in the range of $10\%-12\%$, when the plane $\mu-T$ is scaled by the respective pseudo-critical temperatures for each volume. Furthermore, the critical lines for each set of parameters lie close to the curve of the equilibrium case. Interestingly, the named results are quantitatively the same when $q$ varies in the interval $0.8\leq q\leq 1.2$, which indicates that in this approximation, the chiral symmetry restoration is robust against the thermal fluctuations. Still, the order of the transition may be affected by the size of the system. 

{ Finally, by considering a series expansion in powers of $\beta\hat{H}$, we compared different distribution functions and their corresponding modified Boltzmann factors. In particular, we analyzed the Gamma, $F$, and log-normal distributions. The findings relate that for those super-statistics prescriptions, the effects of the temperature fluctuations and volume remains, i.e., up to the presented approximation, the critical values of $\mu$ and $T$ for the chiral symmetry restoration are the same when $q$ varies. Moreover, the CEP's location moves to higher $\mu$ and lower $T$ when the volume is decreased.}

\section*{Acknowledgement}
We would like to thank Prof. A. Ayala for his useful comments and suggestions about this work. 
\bibliography{bibCEPSE.bib}
\end{document}